\documentclass[preprint]{aastex62}

\shorttitle{NGC6946-BH1}
\shortauthors{R. Humphreys}

\begin{document}

\title{Comments on the Progenitor of NGC6946-BH1 }

\correspondingauthor{Roberta Humphreys}
\email{roberta@umn.edu}

\author{Roberta M. Humphreys}
\affiliation{Minnesota Institute for Astrophysics\\
University of Minnesota\\
Minneapolis, MN, 55455 USA}






\section{} 

\citet{Gerke}  and \citet{Adams} have proposed that the rapid decline of a 
transient in NGC 6946 was a failed supernova due to  the core-collapse of a red 
supergiant to a black hole.  Here  I suggest that the progenitor was instead an 
intermediate temperature star, a yellow hypergiant \citep{deJager}, 
similar to objects like IRC~+10420 \citep{Oud,RMH97,RMH02}  and Var A in M33 
\citep{HS,RMH87,RMH06}. It  was 
on an evolutionary track back to warmer
temperatures when the collapse occurred \citep{RMH13,Gordon}. The properties of  yellow 
hypergiants, including high mass 
loss episodes, strong stellar winds, and dusty circumstellar ejecta, are 
relevant to understanding some of the features in the 
pre-collapse light curve of the progenitor.

\vspace{2mm} 
Figure 1 shows the observed colors and magnitudes corrected for foreground 
extinction. 
\citet{Adams} reject any additional local or environmental extinction in 
NGC 6946. Due to its low Galactic latitude, however, the foreground extinction, 
(A$_{V}$ of 0.94 mag) is rather high.
They use the same in their models.  The broad-band colors of the
progenitor from 1999 up to 2008 are those of an
intermediate temperature star. The most complete set of UBVR photometry is from
2005.5. For example, $(B-V)_{0}$ of 0.8 mag, suggests an 
early-G-type supergiant with a surface temperature of about 5000K. Similarly, 
the other colors are consistent with a late F to early G-type supergiant with 
temperatures of 5000 to  6000K.
Two of their progenitor models (Table 2 in \citet{Adams}) for this 
period indicate temperatures for the star of 6000 to 7000K. 

\begin{figure}
\figurenum{1}
\epsscale{0.7}
\plotone{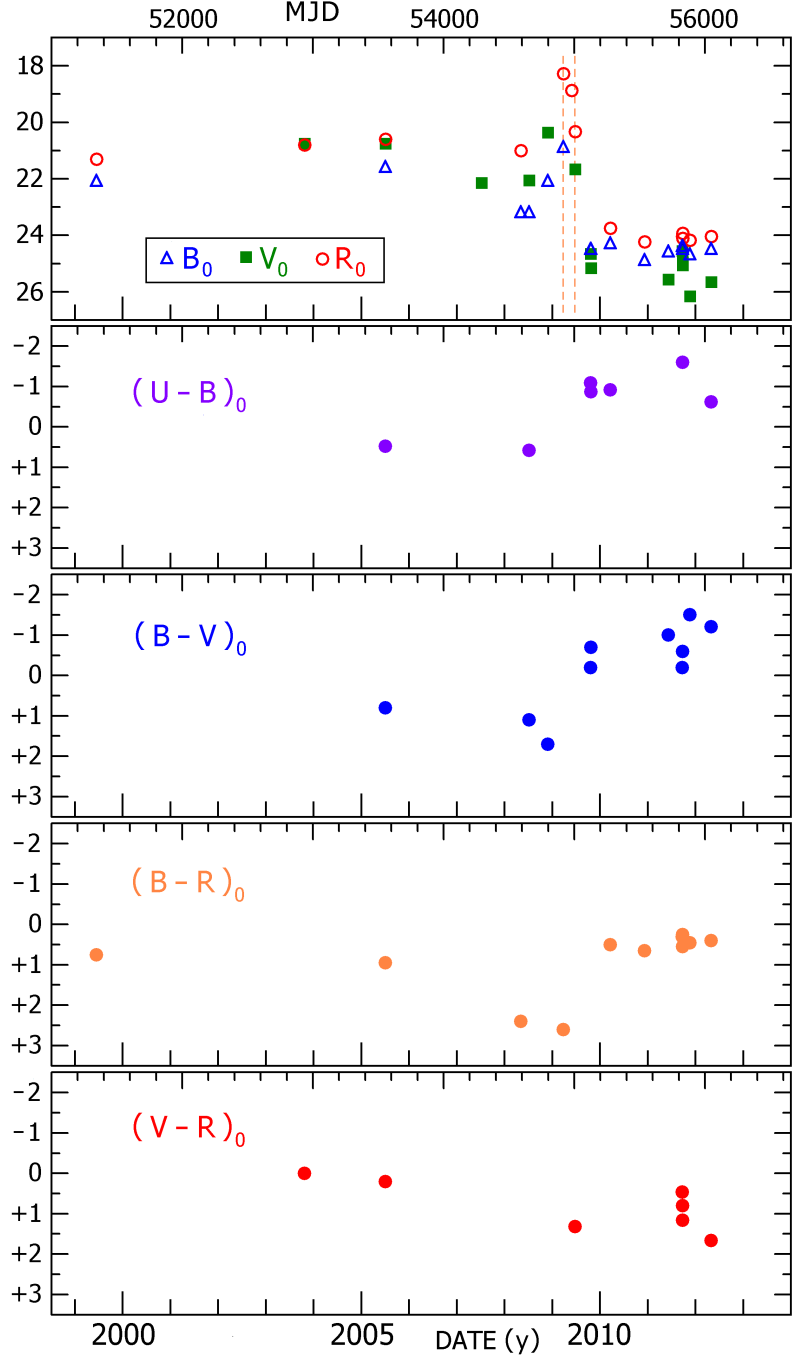}
\caption{The extinction-corrected magnitudes and colors. The two dashed lines 
mark the approximate times  of  maximum brightness (2009.23) and the rapid 
collapse (2009.47). Data from the Appendix in \citet{Adams}.}
\end{figure}

\vspace{2mm} 

The SED for 2005.5 (Figure 2), fit with a 
simple Planck curve, yields a temperature of 5000K.  
Spitzer IRAC 3.6 and 4.5$\mu$m observations from the same period show a 
significant infrared excess due to circumstellar dust and/or free-free 
emission.  The same models \citep{Adams} yield dust temperatures of 
$\sim$ 1700K  which is unusually high. The typical condensation temperature for dust around evolved stars is $\approx$ 1000K \citep{Suh}. The near-infrared flux in the SED is consistent with a strong free-free component, another 
characteristic of the yellow hypergiants with high mass loss and strong winds. 
This does not rule out a contribution from dust at these and longer wavelengths. Integrating over the  SED and the infrared  component out to 
4.5$\mu$m yields a total luminosity  of 2$\times 10^{5}L_{\odot}$. 
 Thus, I suggest a 5000 to 6000K yellow hypergiant for the progenitor. 
 This is different from a red 
 supergiant, an M-type star on the far red side of the HR Diagram. 
 Temperatures for RSGs are $<$ 4000K, typically about 3600K \citep{Lev}.

\begin{figure}
\figurenum{2}
\epsscale{0.8}
\plotone{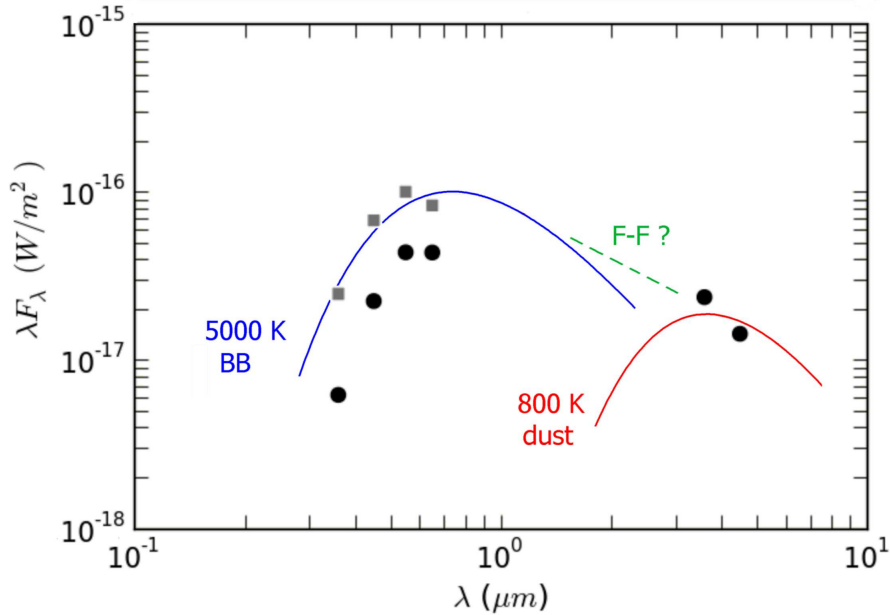}
\caption{The SED from 2005.5. The observed photometry is shown as filled circles and the extinction-corrected fluxes as gray boxes with a 5000K Planck curve. 
The SED demonstrates that free-free emission from the object's wind is 
consistent with the near-infrared fluxes.}
\end{figure}

\vspace{2mm}

A curious feature of the pre-collapse light curve is the decrease in brightness 
beginning in mid-2008 observed at all of the optical wavelengths during which 
the star also got significantly redder; indeed, with the colors of a true 
red supergiant right before the brief outburst, $(B-V)_{0} = 1.7$ (Figure 1). Its SED at this time is 
shown in Figure 3. Many of the yellow hypergiants are 
observed to have high mass loss episodes as recorded in the complex 
circumstellar ejecta of IRC~+10420 \citep{RMH97}, the brief dense wind 
episodes in 
$\rho$ Cas \citep{Lobel} during which it gets redder with the appearance of TiO
bands, and most interestingly in the light curve for Var A in M33 \citep{HS}. 
Var A had a high
mass loss episode, but instead of getting brighter it declined several 
magnitudes due to a  shift in its bolometric correction from the formation a cool dense 
wind with TiO bands and circumstellar dust \citep{RMH87,RMH06}. 
Its cool dense wind, high mass loss state lasted about 50 years.

\vspace{2mm} 

The progenitor’s surface may have been experiencing an event similar to 
Var A when the core collapse occurred. During this same time the infrared shows 
a slow increase in
flux very likely  due to the enhanced mass loss and possible dust formation. 
\citet{Adams} attribute the subsequent brief brightening to ejection of the 
supergiant’s envelope.
Post-collapse photometry shows very blue colors (Figure 1). But it is
uncertain how realistic the published photometry and colors are. Some of the 
reported very faint magnitudes may
be equivalent to non-detections. HST and Spitzer visual and near-infrared 
magnitudes from 2015.7 present a different picture. The corresponding SED 
(Figure 3)  reveals a source 
with most of the radiation arising from the 3 to 5 micron region. It is unclear 
if this is remnant dust from the
progenitor’s circumstellar ejecta or new dust formed from the outburst. 

\begin{figure}
\figurenum{3}
\epsscale{1.1}
\plotone{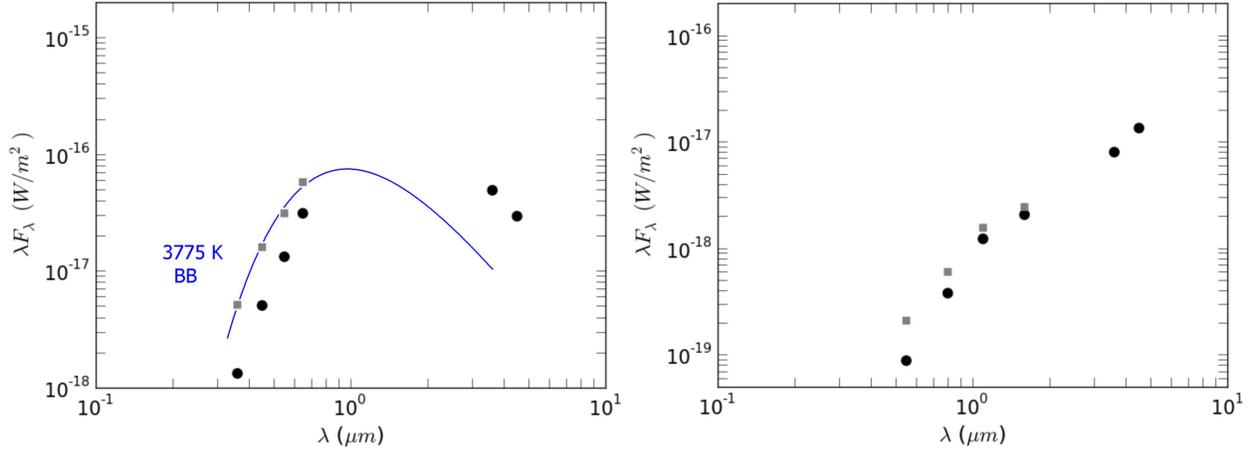}
\caption{Left: The SED from 2008.5 with a 3775K Planck curve. Right: The post collapse SED from 2015.7 from optical and near-infrared measurements from HST and Spitzer \citep{Adams}.}
\end{figure}

\vspace{2mm}

In summary, the pre-outburst and pre-collapse observations of NGC 6946-BH1 show
that the progenitor was a yellow hypergiant probably experiencing high mass loss
before the collapse. Some might question what difference does it make, 
yellow supergiant or red supergiant. This star was obviously highly evolved 
and was therefore in a post-RSG state immediately before the core collapse. Not 
only was its surface hotter, its radius was much smaller, and its atmosphere or envelope was denser; all factors that will make a difference in the models 
for the outburst and collapse.

\end{document}